\documentclass[10pt,preprint]{aastex}
\usepackage{color}

\newcommand{\av}{$A_V$}

\newcommand{\etal}{et~al.}

\newcommand{\ks}{$K_{\rm s}$}

\newcommand{\mum}{$\mu$m}

\begin{document}

\title{New Young Star Candidates in the Taurus-Auriga Region as
Selected from WISE}

\slugcomment{Version from \today}

\author{L.\ M.\ Rebull\altaffilmark{1}, 
X.\ P.\ Koenig\altaffilmark{2}, 
D.\ L.\ Padgett\altaffilmark{1}, 
S.\ Terebey\altaffilmark{3}, 
P.\ M.\ McGehee\altaffilmark{4},  
L.\ A.\ Hillenbrand\altaffilmark{5}, 
G.\ Knapp\altaffilmark{6}, 
D.\ Leisawitz\altaffilmark{2}, 
W.\ Liu\altaffilmark{4},  
A.\ Noriega-Crespo\altaffilmark{1}, 
M.\ Ressler\altaffilmark{7}, 
K.\ R.\ Stapelfeldt\altaffilmark{7},
S.\ Fajardo-Acosta\altaffilmark{4},
A.\ Mainzer\altaffilmark{7}
  }

\altaffiltext{1}{Spitzer Science Center/Caltech, M/S 220-6, 1200
E.\ California Blvd., Pasadena, CA  91125
(luisa.rebull@jpl.nasa.gov)}
\altaffiltext{2}{Goddard Space Flight Center, Greenbelt, MD}
\altaffiltext{3}{Cal State University, Los Angeles}
\altaffiltext{4}{Infrared Processing and Analysis Center}
\altaffiltext{5}{California Institute of Technology}
\altaffiltext{6}{Princeton University}
\altaffiltext{7}{Jet Propulsion Laboratory}

\begin{abstract}

The Taurus Molecular Cloud subtends a large solid angle on the sky, in
excess of 250 square degrees. The search for legitimate Taurus members
to date has been limited by sky coverage as well as the challenge of
distinguishing members from field interlopers. The Wide-field Infrared
Survey Explorer (WISE) has recently observed the entire sky, and we
take advantage of the opportunity to search for young stellar object
(YSO) candidate Taurus members from a $\sim$260 square degree region
designed to encompass previously-identified Taurus members.  We use
near- and mid-infrared colors to select objects with apparent infrared
excesses and incorporate other catalogs of ancillary data to present:
a list of rediscovered Taurus YSOs with infrared excesses (taken to be
due to circumstellar disks), a list of rejected YSO candidates
(largely galaxies), and a list of 94 surviving candidate new YSO-like
Taurus members.  There is likely to be contamination lingering in this
candidate list, and follow-up spectra are warranted.

\end{abstract}


\section{Introduction}
\label{sec:intro}

In recent years, there have been many infrared (IR) studies of nearby
star-forming regions with the Spitzer Space Telescope (Werner \etal\
2004).  Since most if not all low-mass stars form with circumstellar
accretion disks, they have IR excesses for as long as the dusty
circumstellar material survives (e.g., Hernandez \etal\ 2008). Stars
with large IR excesses are relatively easily distinguished from stars
without such excesses, while stars with more modest excesses require
more accurate knowledge of the underlying star and the intervening
extinction. Surveys with Spitzer have proven very good at finding new
young stars in star-forming regions, some located surprisingly far
from the traditional locations of star formation based on CO gas or
IRAS dust maps. However, even though Spitzer is able to survey large
areas relatively quickly, it still is a pointed observatory, and
cannot easily conduct a truly wide-field  survey.  In very close
star-forming regions such as Taurus, the large solid angle subtended
by the association challenges the search for new members, rendering
the samples incomplete, particularly for a distributed population. 
See, however, Rebull \etal\ (2010).

We conducted a $\sim$44 square degree survey of Taurus with Spitzer,
using 3.6, 4.5, 5.8, 8, 24, 70, and 160 \mum\ (Padgett \etal\ in prep,
Padgett \etal\ 2008a, G\"udel \etal\ 2007).  In Rebull \etal\ (2010),
we reported on our search for new young stellar objects (YSOs) in
Taurus. We used a primarily Spitzer-driven color selection, but took
advantage of considerable ancillary data amassed in the service of a
multi-wavelength search for new young stars. We found that any solely
near- and mid-infrared color selection was filled with contamination
from galaxies and asymptotic giant branch (AGB) stars, and that use of
ancillary data was crucial to establishing a list of high-quality new
members of Taurus. We now use our experience with this prior data set
to inform our selection using Wide-field Infrared Survey Explorer
(WISE; Wright \etal\ 2010) data.

WISE conducted an all-sky survey in 2009-2011. These data are
well-suited to studying nearby star-forming clouds such as Taurus,
where members are likely to be bright because of their proximity to us
as well as their youth, and well within the regime of good
signal-to-noise ratio (SNR) WISE photometry.  Furthermore, the all-sky
nature of the survey allows us to look away from the more traditional
and well-studied clustered regions and include some young stars in the
direction of Taurus (not necessarily Taurus members) that were never
studied with Spitzer. 

WISE surveyed the entire sky, but the depth of coverage is non-uniform
and generally greater at higher ecliptic latitudes (Taurus is in the
ecliptic plane). The depth of coverage in the Taurus region is
somewhat degraded relative to regions of comparable ecliptic latitude
due to Moon avoidance maneuvers made during the mission.
Most of the Taurus region is observed to a depth of about 0.08, 0.11,
0.8, and 6 mJy in the four bands (5$\sigma$, given 8 visits), and we
further restrict the depth with signal-to-noise ratio (SNR) cuts; see
below. Our original Taurus Spitzer Survey went to 0.06, 0.06, 0.14,
and 1.1 mJy for IRAC-1 (3.6 \mum), IRAC-2 (4.5 \mum), IRAC-4 (8 \mum),
and MIPS-24 (24 \mum), respectively. However, both surveys should
easily detect legitimate Taurus members, since the cloud is only 140 pc
away.

In this paper, we select new candidate Taurus members with infrared
excesses using WISE colors using the method established in Koenig
\etal\ (2011).  We compare the list of selected objects to catalogs we
have assembled from our prior work and updated. We report three lists
-- recovered young stars, rejected objects, and candidate new Taurus
members.  The observations and basic data handling are described in 
\S\ref{sec:obs}, \S\ref{sec:findysos} describes how we identified our
YSO candidates, \S\ref{sec:properties} presents some global properties
of the ensemble of YSO candidates, with a special focus on estimating
the degree of contamination, and finally we summarize in section
\ref{sec:concl}.

\section{Observations, Data Reduction, and Ancillary Data}
\label{sec:obs}

\begin{figure*}[tbp]
\epsscale{1}
\plotone{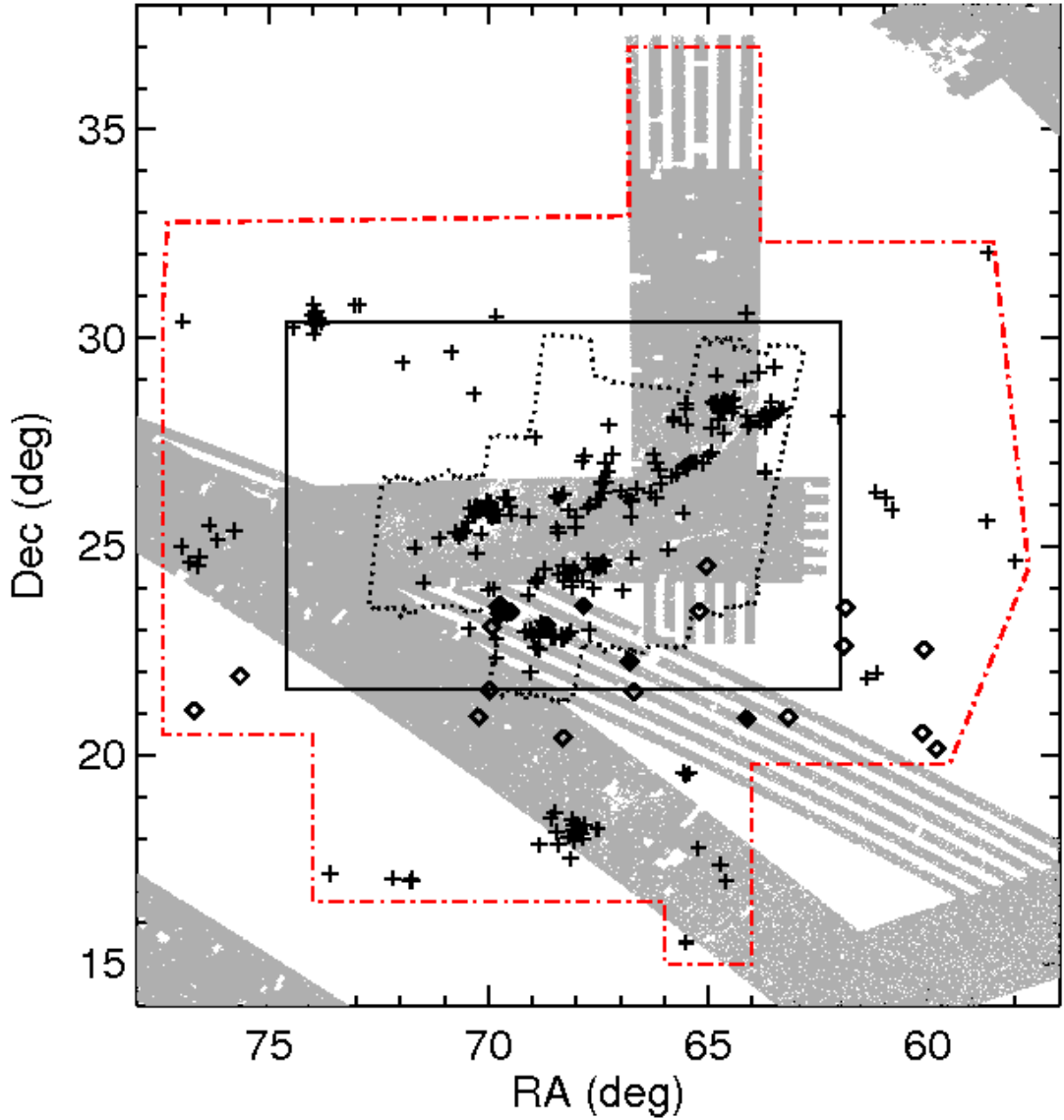}
\caption{Location in the sky of the various surveys discussed here.
Solid black box: boundary of Goldsmith \etal\ (2008) CO(1-0) survey;
grey-colored regions: SDSS coverage; smaller irregular black dotted
polygon: coverage of Spitzer Taurus Survey (Padgett \etal\ in prep,
Padgett \etal\ 2008a, G\"udel \etal\ 2007); red dash-dot line: boundary
of polygon extracted from WISE catalog; $+$ symbols: Taurus members
from Kenyon \etal\ (2008); diamonds: proposed Taurus members from
Slesnick \etal\ (2006). }
\label{fig:where}
\end{figure*}

\begin{figure*}[tbp]
\epsscale{1}
\plotone{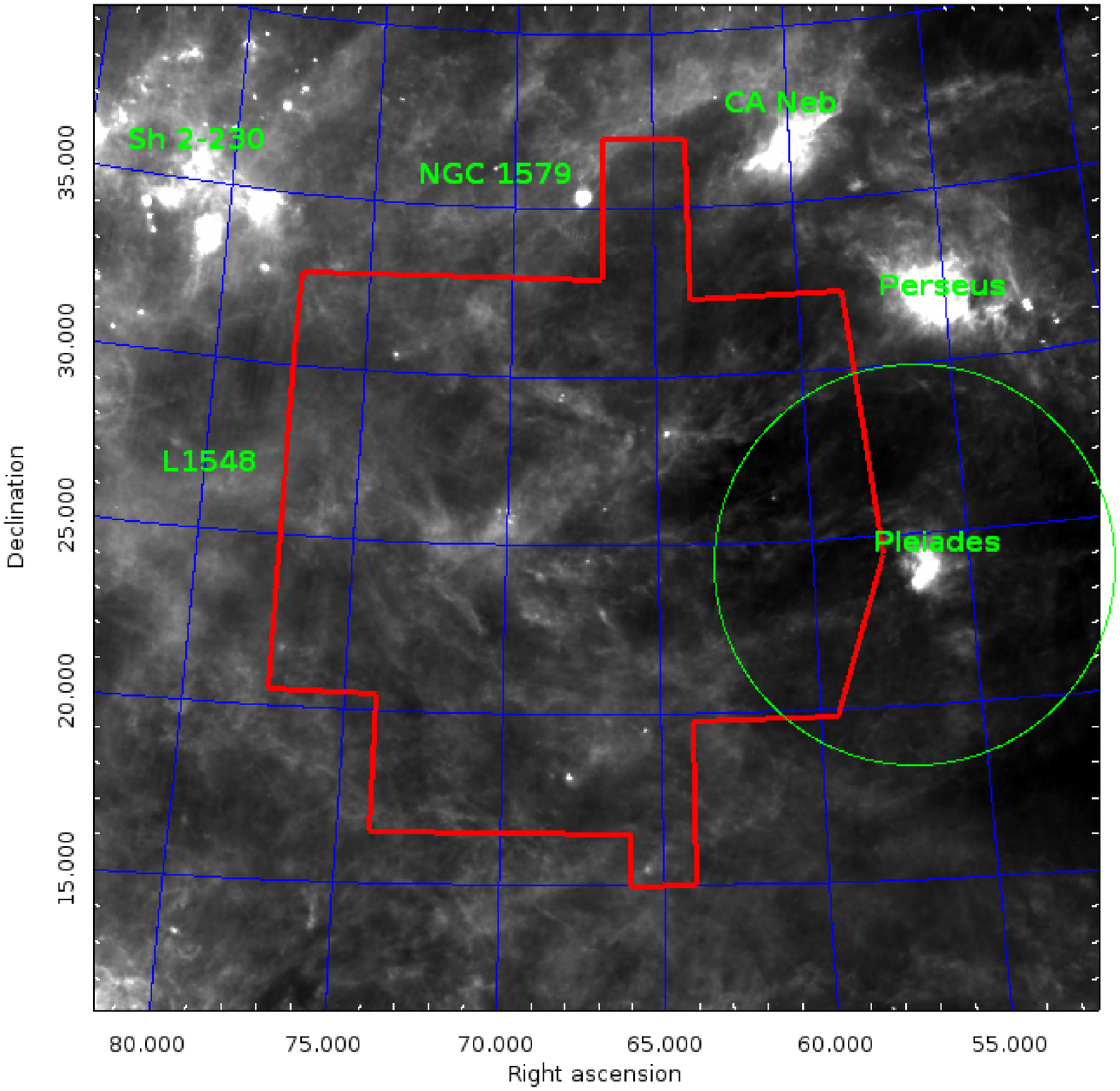}
\caption{Overlaid on an IRAS 100 \mum\ map, the location in the sky of
our survey contour (red polygon), other nearby star-forming regions of
interest, and the tidal radius of the Pleiades (green circle). L1548
will be discussed by Liu \etal, in preparation.  Some of our objects
are within the tidal radius of the Pleiades and thus perhaps could be
members of the Pleiades rather than Taurus. NGC 1579 is centered on
the bright spot. }
\label{fig:whereiras100}
\end{figure*}

In the context of Rebull \etal\ (2010), we assembled a substantial
multi-wavelength database, spanning Sloan $u$ through Spitzer/MIPS 160
\mum\  (with some X-rays) for point sources throughout the Taurus
region. We note that not every source has photometry at all bands due
to variations in depth and spatial coverage among the surveys
involved.   We use that catalog as the core for our analysis here,
updating it with confirmed Taurus members from, e.g., Kenyon \etal\
(2008) and Luhman \etal\ (2010) outside of our original Spitzer map. 
We have also searched SIMBAD (and literature references therein) for
known galaxies and other contaminants such as planetary nebulae (PNe)
in this viscinity and considered these identifications in our analysis
of the WISE photometry.  

Our Taurus Spitzer Survey spanned $\sim$44 square degrees; see
Figure~\ref{fig:where}. Notably, two of the surveys we assembled at
other wavelengths extend well beyond the region we mapped with Spitzer
-- the CO(1-0) radio map (Goldsmith \etal\ 2008) and two Sloan Digital
Sky Survey (SDSS) stripes. Goldsmith \etal\ (2008) mapped $\sim$100
square degrees; see Figure~\ref{fig:where}. The SDSS  (Finkbeiner
\etal\ 2004; Padmanabhan \etal\ 2008) initially mapped the Taurus
region in two perpendicular strips covering $\sim$48 square degrees,
overlapping in part with the Goldsmith \etal\ (2008) and our Spitzer
maps, but extending further east-west and substantially further
north.  Additional SDSS strips substantially increased the area
observed; see Figure~\ref{fig:where}.  We note that the unfilled 
stripe that crosses the region diagonally is one of the photometric
calibration stripes a la Padmanabhan \etal\ (2008).  Further motivating
consideration of a wider area, the  XMM-Newton Extended Survey of the
Taurus Molecular Cloud (XEST) program (G\"udel \etal\ 2007 and
references therein) mapped $\sim$5 square degrees, most but not all of
which was covered by our Spitzer maps.  Finally, as seen in
Figure~\ref{fig:where}, Kenyon \etal\ (2008) includes a list of known
Taurus members, many of which are well beyond the bounds of our
original Spitzer survey, and Slesnick \etal\ (2006) report on several
more candidate Taurus members. 

We define a polygon in which we searched for YSO candidates using WISE
that encompasses all of the members from Kenyon \etal\ (2008), a few
of the potential members from Slesnick \etal\ (2006), and all of the
Sloan coverage to the north of our previous Spitzer survey.  The
RA/Dec vertices of this polygon are, in degrees: 77.4 31, 77.4 20.5,
74 20.5, 74 16.5, 66 16.5, 66 15, 64 15, 64 19.8, 59.5 19.8, 57.7
24.5, 58.5 32.3, 63.8 32.3, 63.8 37, 66.8 37, 66.81  32.93, 77.30
32.77. The total area we have considered is $\sim$260 square degrees.

This area is so large that we need to consider other 
star-forming regions very nearby (in projection) to Taurus. 
Figure~\ref{fig:whereiras100} shows our polygon on an IRAS 100 \mum\
image of the region.  NGC 1579 is the bright knot in the central north
of the image; this is thought to be 700 pc away (Herbig \etal\ 2004),
too far  to be part of Taurus at 140 pc. Lynds 1548 and environs will
be discused in a forthcoming paper by Liu et al. Our polygon
intersects with the tidal radius of the Pleiades ($\sim$6 degrees;
Adams \etal\ 2001). Young objects identified within our polygon and within
$\sim$6 degrees of the Pleiades could belong to the Pleiades and not
Taurus.

WISE data acquisition and reduction are discussed in Wright \etal\
(2010), Jarrett \etal\ (2011), and in the Explanatory Supplement to
the WISE Preliminary Data Release Products.  There are four WISE
bands, with central wavelengths at 3.4, 4.6, 12, and 22  \mum, and a
spatial resolution of 6$\arcsec$ (12$\arcsec$ at 22 \mum).   The four
bands are often referred to as W1, W2, W3, \& W4. We rejected any
source with contamination and confusion flags (``cc\_flags'' in the
catalog) that include any of the characters ``DHOP'' in its
four-character string as likely contamination or confusion
artifacts\footnote{``D'' denotes diffraction spike, ``H'' indicates
that the source is spurious as a result of, or contaminated by, the
scattered light halo surrounding a nearby bright source, ``O'' denotes
optical ghost, and ``P'' denotes persistence artifact.}. This process
resulted in a catalog containing about 2.38 million sources.   The
WISE catalog reports signal-to-noise ratios.  We further restricted
ourselves to those measurements with SNR $>$ 7 in all four bands,
drastically shrinking the catalog to $\sim$7000 sources.  Our initial
attempts  limited the analysis to those measurements with SNR $>$ 7 in
just the  first three bands, but given the expected brightness of
Taurus members, plus the contamination rate we expect based on our
experience with Spitzer in Taurus, we opted to enforce the SNR cut in
W4 as well so as to limit the contamination in our list of candidate
YSOs.

The WISE catalog reports flux measurements in magnitudes with errors
in magnitudes, and the source selection in this paper was based on these
magnitudes.  The zero-points we used to convert between magnitudes and
flux densities in the spectral energy distribution (SED) 
plots included in the online-only appendix are
309.54, 171.79, 31.676, and 8.3635 Jy for the four channels, respectively,
based on the zeropoints in Wright \etal\ (2010) and the flux correction
factors for a spectrum with constant $F_{\nu}$.  These zeropoints are also
used by Jarrett \etal\ (2011).

The WISE bands are merged among themselves and to the Two-Micron All-Sky
Survey (2MASS; Skrutskie \etal\ 2006) in the data products provided by the
WISE archive housed at the Infrared Science Archive (IRSA).   We
merged this catalog to our already-assembled catalog of ancillary data
via a strictly position-based search with a 1$\arcsec$ radius. 
As noted in Section VI.5.f of the WISE Explanatory Supplement, the
WISE catalog reports positions of sources brighter than W1$<$ 13.0 mag
to 0.2 arcsec (1 sigma) relative to 2MASS positions.  The positions
reported for fainter sources may suffer from a systematic error up to
1$\arcsec$ in Declination due to a pipeline coding error.

\section{Identifying the YSO Candidates}
\label{sec:findysos}

The process of identifying the YSO candidates is a multi-step process,
beginning with color cuts and progressing through ancillary data,
including manual checks of spectral energy distributions (SEDs) and
images. We now describe the process that we used.

\subsection{Initial cut}

We started with our extracted WISE catalog over the $\sim$260 square
degree polygon specified above, and applied the color cuts discussed
in Koenig \etal\ (2011).  In summary, there are a series of cuts in
multiple color spaces based on the location of previously identified
YSOs and galaxies. These are intended statistically to weed out most 
contaminants and find most YSOs. No color cuts can perform this task
flawlessly, though many have been discussed in the literature in the
context of Spitzer observations (e.g., Allen \etal\ 2004, Padgett
\etal\ 2008b, Rebull \etal\ 2007, Harvey \etal\ 2007, Gutermuth \etal\
2008, 2009, Rebull \etal\ 2010, 2011). As discussed in Rebull \etal\
(2010), especially over very large fields like Taurus, where the
molecular cloud does not block out most background sources and where
the survey area is big enough that the chances of obtaining more
unusual objects are greater, the contamination rate for any color
selection is expected to be relatively large, and ancillary data are
crucial for culling the list to high-quality candidates. 

The color cuts described in Koenig \etal\ (2011) are inspired by 
Gutermuth \etal\ (2008, 2009) and Rebull \etal\ (2010) and applied
using WISE+2MASS colors. For their full selection process, Koenig
\etal\ (2011) estimate a contamination rate for `typical' star forming
regions of about 2.4 objects resembling Class Is, 3.8 objects
resembling Class IIs and 1.8 objects resembling transition disks per
square degree. At this rate, with our $\sim$260 square degree map, we
expect $\sim$620, $\sim$990, and $\sim$470, respectively, for a rough
total of $\sim$2000 contaminants per square degree. As a check, we
recalculated these contamination rates for a $\sim$10 square degree
patch at the north and south equatorial poles, and obtained values
between $\sim$1500 and $\sim$1600 contaminants per square degree.
However, these values are sensitive to relative depths of WISE
coverage (deeper at the poles than the ecliptic), any bright extended
emission in the image (present here but less so in the extragalactic
fields), and natural variability in the space density of stars and
galaxies.

We initially applied the Koenig \etal\ method to the WISE catalog
where the SNR is at least 7 in channels 1, 2, and 3,  but not
necessarily 4. We thus obtained $\sim$1760 YSO candidates. Given the
contamination rates expected above, we thus anticipate a high fraction
of contaminants in this list.  As mentioned above, given the expected
brightness of Taurus members, plus the contamination rate we expect
based on our experience with Spitzer in Taurus, we chose to impose a
SNR cut in W4 as well so as to limit the contamination in our list of
candidate YSOs. Imposing our additional requirement that the SNR is at
least 7 in all four WISE channels results in a pool of 1014 potential
YSOs.  

\subsection{Ancillary data}


We know from our Spitzer search for new members of Taurus that
ancillary data are very important for weeding the contaminants out
from the list of potential YSO candidates. So, in addition to our
catalog above, we have included the information from ancillary data in
our assembly of our final YSO candidate list in an effort to limit the
contamination.

We have matched our WISE catalog to the full SDSS catalog (from DR8)
in this region (which includes extended source information from the
images) and the 2MASS Extended Source Catalog. Objects that are
extended are likely to be (though are not guaranteed to be) galaxies;
see discussion in Rebull \etal\ (2010).  There are also $\sim$11,500
SDSS spectra in the SDSS stripes; just 27 of the sources on our list
of potential YSOs find matches with SDSS spectra.

We merged to the Akari 9 and 18 \mum\ IRC all-sky point source catalog
(Ishihara \etal\ 2010). There are $\sim$3000 sources in this region,
$\sim$100 of which match to the WISE-selected sources, $\sim$80\% of
which are for previously-known Taurus members.  

Based on our experience with our Taurus Spitzer Survey, we know that
the YSOs are generally though not exclusively found in regions of 
high \av.  Thus, we expect that an estimate of \av\ towards our new
candidates in this larger region will also be useful in weeding out
contaminants. Froebrich \etal\ (2007) report on a large, 127 $\times$
63 square degree  extinction map based on 2MASS data.  We used this
map to estimate \av\ towards our list of candidates. This map is
calculated to a $\sim$4$\arcmin$ resolution, and we used a 3$\times$3
pixel ($6\arcmin\times6\arcmin$) median calculated about the position
of each source to estimate the \av.

\subsection{Previously Identified YSOs}
\label{sec:knownysos}

For the 1014 potential YSOs selected from WISE+2MASS color and
magnitude cuts imposed on the 2.38 million sources with good SNR
photometry, 196 of them have matches to previously-identified stars
with indications of youth and/or infrared excesses, in the direction
of Taurus.  Table~\ref{tab:knownysos} lists these objects and their
WISE measurements. Most of the objects in Table~\ref{tab:knownysos}
are previously identified explicitly as Taurus members in G\"udel
\etal\ (2007) and references therein, Kenyon \etal\ (2008), Rebull
\etal\ (2010), and/or Luhman \etal\ (2010). Eighteen of the objects in
Table~\ref{tab:knownysos} are listed as unconfirmed candidates in
Rebull \etal\ (2010); these are identified as such in the ``notes''
column of Table~\ref{tab:knownysos}.  (As discussed in Rebull \etal\
2010, for these objects, we could not find unambiguous spectroscopic
indications of youth -- such as the H$\alpha$ emission line was low or
absent -- so additional data are needed to confirm or refute these
objects as Taurus members.  These objects should not be regarded as
Taurus members with the same confidence as, e.g., DG Tau, but they are
not necessarily clearly field interlopers either.)  The candidates
from Slesnick \etal\ (2006) would have been similarly indicated,
except they are largely not recovered, save for three objects commonly
taken as Taurus members (SCHJ0429595+2433080=CFHT-20,
SCHJ0438586+2336352=J0438586+2336352, and
SCHJ0439016+2336030=J0439016+2336030).   Eight of the objects in
Table~\ref{tab:knownysos} are not traditionally identified as Taurus
members, but we recover them as having IR excess, and they appear in
the literature as having some indications of youth and with
appropriate proper motions for Taurus members (2MASS
J04360131+1726120, 2MASS J04324107+1809239, HD 285893, HBHA 3214-06,
GZ Aur, BS Tau, IRAS 05020+2518, HO Aur). These are also indicated in
Table~\ref{tab:knownysos}, in the ``notes'' column.

Note that this table is not meant to be a complete list of Taurus
members detected with WISE, but instead a list of all objects in the
direction of Taurus, identified as young, detected in WISE with SNR
$>$ 7 in all four bands and identified using the Koenig \etal\ (2011)
method as having an IR excess. Some bona fide members of Taurus are
either too faint or do not have enough color excess to be identified
in this manner.

\subsection{Manual inspection}

For the remaining 818 WISE-identified IR excess objects that have not
been associated previously with young stars, we conducted a
preliminary sorting into ``likely contaminant'' or ``perhaps YSO''
bins. The categorization was based on matches to objects in SIMBAD
(and literature therein), matches to objects we identified as
contaminants in Rebull \etal\ (2010), matches to the 2MASS Extended
Source Catalog, and identification as extended in the SDSS pipeline.
We then generated SEDs using all photometric data in our database for
these objects, and inspected each of them.  Based on our experience,
we then categorized each of the objects as still possible YSO
candidates, or likely extragalactic objects. This classification is
easier when there are more photometric bands; objects for which there
are SDSS, 2MASS, and WISE photometry are more likely appropriately
classified during this process than those without optical photometry
or with just WISE photometry. Therefore, this process may have dropped
viable YSO candidates similar to e.g., MHO-1 or Haro 6-39, which have
very flat SEDs (and moreover MHO-1 has just WISE photometry).  
Statistically the objects we have dropped on this basis are very
likely to be extragalactic. Because this process is not perfect, the
objects dropped as a result of SED shape are identified in the
``notes'' columns of the Tables below specifically so that follow-up
optical photometry (for example) can be obtained.

After this process, there were $\sim$130 objects which we could not
yet reject as YSO candidates. For these objects, we inspected the
images in all four WISE bands, and, if necessary, 2MASS, POSS, and if
possible, SDSS. We then identified objects as likely subject to source
contamination, resolved as a likely galaxy, or still apparently clean,
relatively isolated point soures.

We have thus identified 686 objects that are confirmed or likely
galaxies, 13 foreground or background stars, 1 planetary nebula, and
24 objects that seem to be subject to confusion in the relatively
large WISE beam (and therefore any excess seen in the SED is likely to
be contaminated by the confused source).  There are 94 objects still
surviving in the potential YSO candidates list. The rejected objects
appear in Table~\ref{tab:rejysos}, and the 94 surviving objects appear
in Table~\ref{tab:newysos}. SEDs for the 94 potential YSO objects
appear in the Appendix.

\section{Overall Properties}
\label{sec:properties}

We now attempt to assess the overall contamination as well as identify
very high-likelihood YSOs out of our candidate list using the
ancillary data we have amassed.


\subsection{Projected Location}

\begin{figure*}[tbp]
\epsscale{1}
\plotone{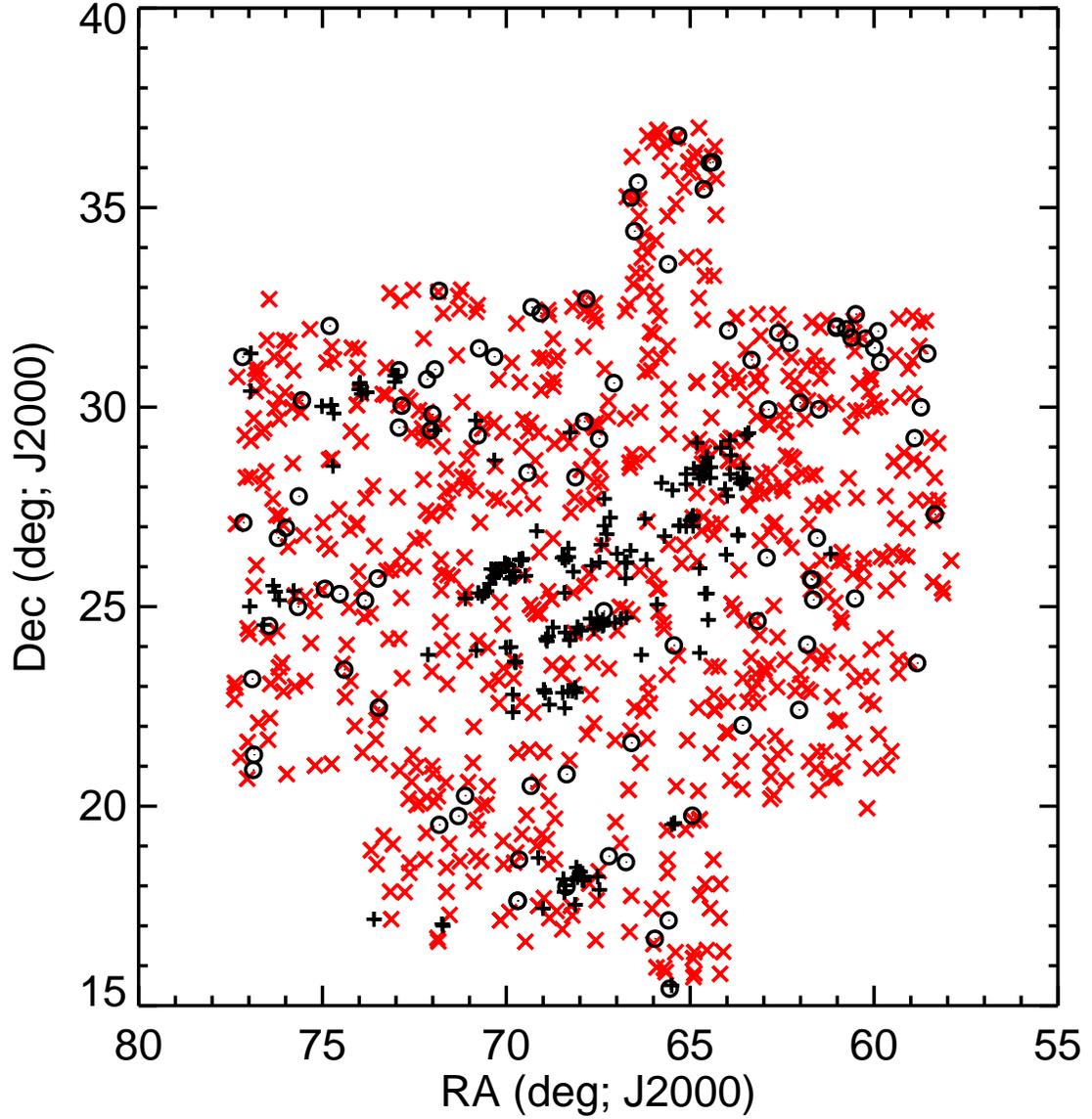}
\caption{Location on the sky of the recovered prevously-identified
YSOs (+), the rejected objects ($\times$), and the best remaining
candidates (circles). The previously-identified YSOs are generally
highly clustered, and the newly identified candidates are less clustered.}
\label{fig:location}
\end{figure*}

Figure~\ref{fig:location} shows the location on the sky of the
recovered previously-identified YSOs, the rejected objects, and the
surviving candidate YSOs. The previously-identified YSOs are generally
highly clustered along the filamentary distribution of gas and dust,
and the new objects are less clustered. This is as expected, since
many searches have focused on the regions already known to contain
young stars, and our goal was to look for new YSOs outside the
canonical groupings of previously-known Taurus members. However, this
could also be an indication of persistent contamination in the
surviving list of YSO candidates.

\begin{figure*}[tbp]
\epsscale{1}
\plotone{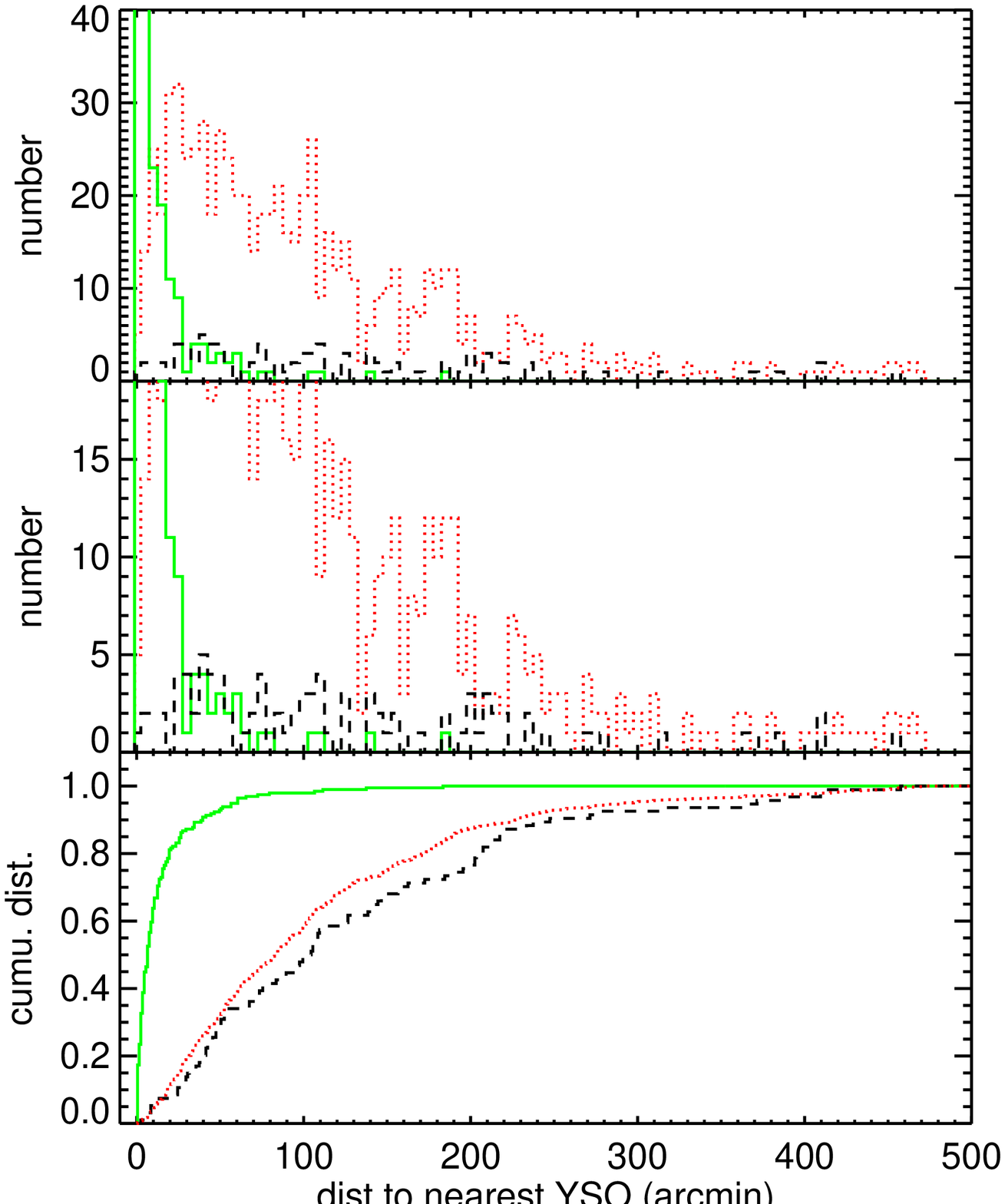}
\caption{Histogram (top two; middle is zoomed-in version of the top)
and cumulative distribution (bottom; fraction of sample with number of
points less than the corresponding $x$ value) of distances in
arcminutes to the nearest previously identified YSO for recovered
known YSOs (green, solid line; distance is to nearest other YSO),
rejected contaminants (red, dotted line), and new YSO candidates
(black, dashed line). }
\label{fig:distance2ysos}
\end{figure*}

Figure~\ref{fig:distance2ysos} is a more quantitative view of the
clustering seen in Figure~\ref{fig:location}, showing a distribution
of separations.  Previously identified YSOs are very close,
on average, to another previously identified YSO.  The contaminants have
a much broader distribution and are further from the known YSOs on
average. The distribution of candidate YSOs is closer to that of the
contaminants than the previously-identified YSOs.

In the context of proximity of objects to other known objects,  as
noted above, some of our objects can be found within the tidal radius
of the Pleiades ($\sim$6 degrees; Adams \etal\ 2001). Objects
identified here that are within $\sim$6 degrees of the Pleiades could
belong to the Pleiades rather than Taurus. One recovered known YSO is
within this radius (IRAS 04016+26102), 104 rejected contaminants, and
12 YSO candidates (J035323.82+271838.3, J035519.09+233501.8,
J035535.05+291319.3, J040204.41+251210.3, J040612.75+264308.0,
J040636.44+251019.1, J040644.43+254018.1, J040651.36+254128.3=V1195
Tau, J040716.73+240257.6, J040809.49+222434.8, J041141.34+261341.3,
and J041240.69+243815.6).

\subsection{Reddening}

\begin{figure*}[tbp]
\epsscale{1}
\plotone{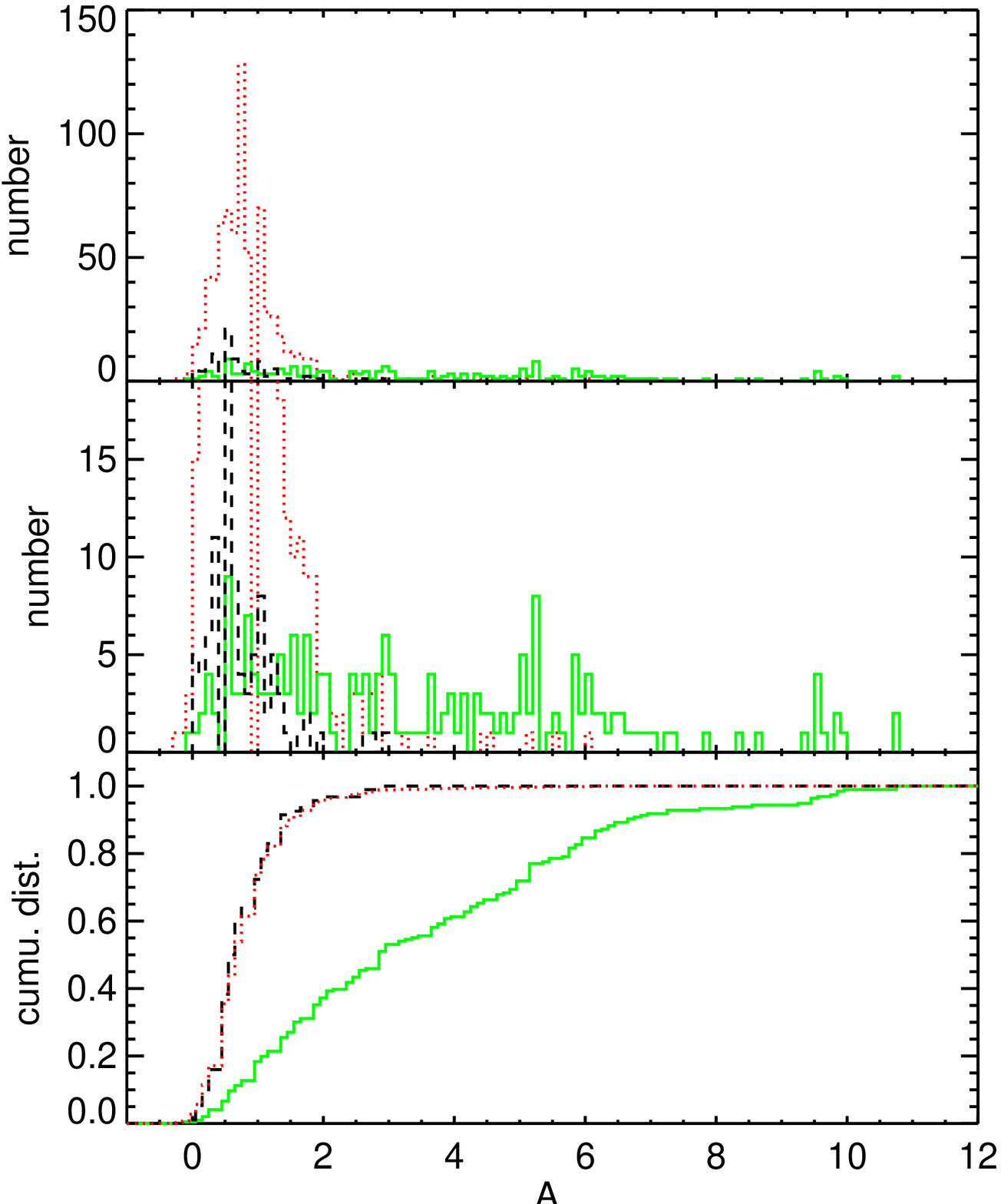}
\caption{Histogram of \av\ values (top two; middle is zoomed-in
version of the top), and cumulative distribution of \av\ values
(bottom), for recovered known YSOs (green, solid line), rejected
contaminants (red, dotted line), and new YSO candidates (black, dashed
line).  The distribution of \av\ values for the candidates is much
closer to the distribution of contaminants than previously known
YSOs.  The \av\ estimates come from a median
$\sim6\arcmin\times6\arcmin$ box centered on the position of the
object.}
\label{fig:av}
\end{figure*}

\begin{figure*}[tbp]
\epsscale{1}
\plottwo{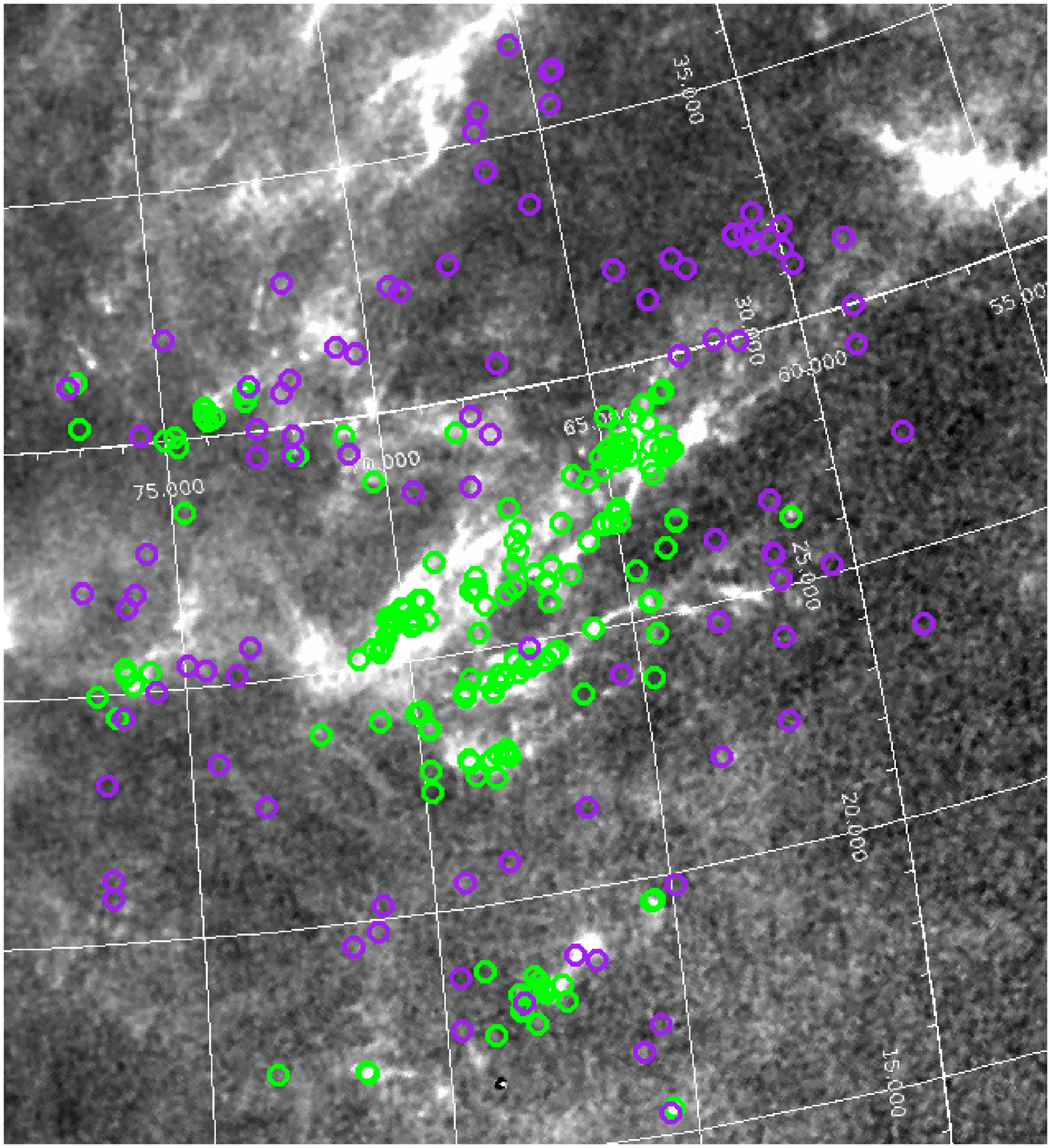}{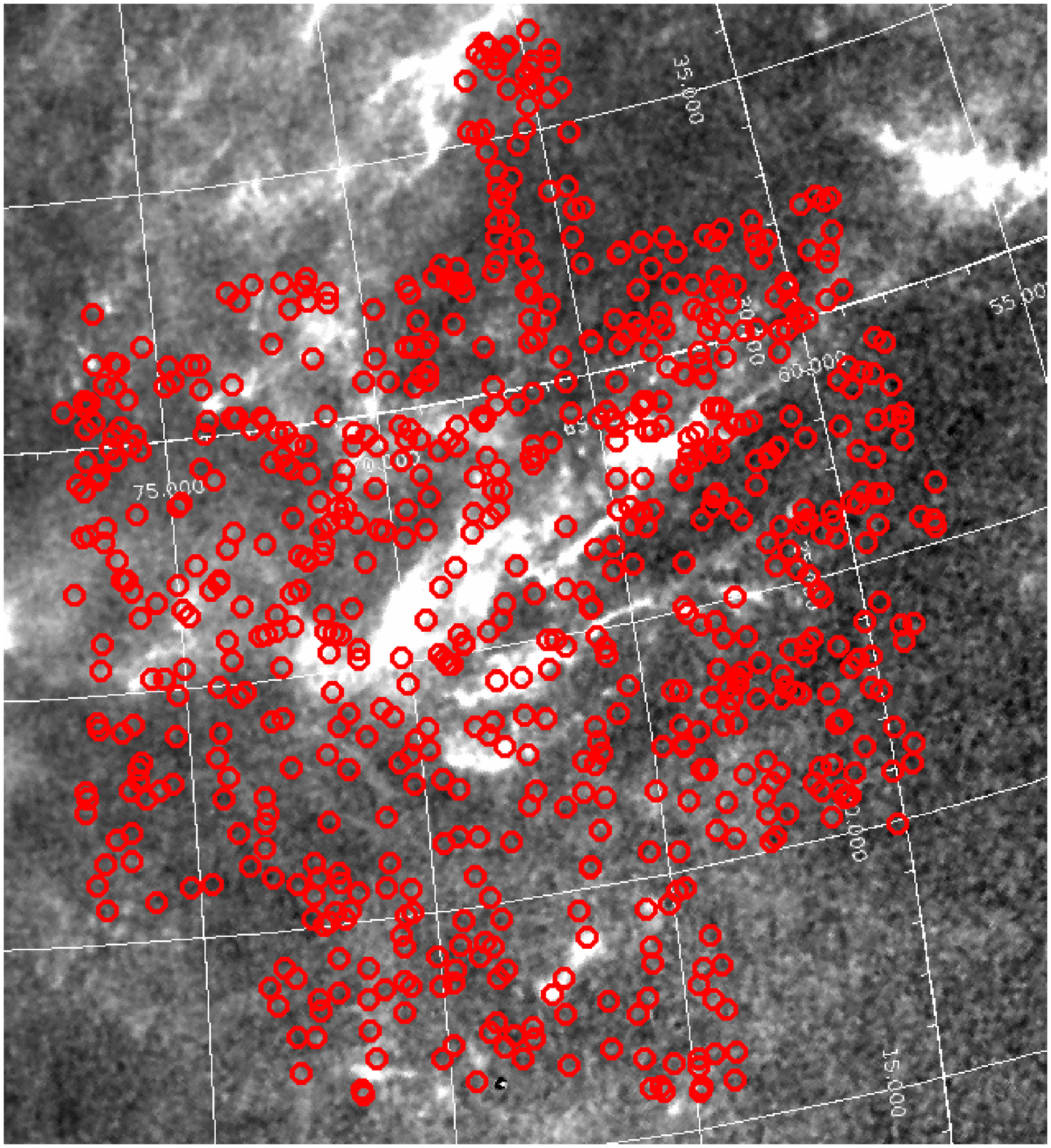}
\caption{Locations of YSO candidates and recovered previously
identified YSOs (left, purple and green, respectively) and rejected
candidates (right) superimposed on an \av\ map constructed from 2MASS
by Froebrich \etal\ (2007) with $\sim$4$\arcmin$ resolution. The
structure seen in the \av\ can be compared to that seen in IRAS-100
\mum\ in Fig.~\ref{fig:whereiras100}.}
\label{fig:avimage}
\end{figure*}

Figure~\ref{fig:av} is another look at this same issue of
contamination in the candidate list, this time through the lens of
\av. The previously known YSOs are generally found in regions of high
\av, and background galaxies are found in regions of low \av.  As
discussed above, we obtained a coarse estimate of \av\ based on a
$\sim6\arcmin\times6\arcmin$ box centered on the position of the
object in a map constructed from 2MASS star counts (Froebrich \etal\
2007).   Figure~\ref{fig:av} shows that the distribution of \av\ for
the candidates is considerably closer to the distribution of \av\ for
the contaminants rather than the previously-known YSOs. The largest
\av\ found for the recovered YSOs is 10.8 mag, and the largest \av\
for the candidates is only 3 mag (J042850.54+184436.1), with two
others having \av$>$2 (J042213.75+152529.9 -- \av=2.6,
J045141.49+305519.5 -- \av=2.7).  These may be among the highest
likelihood YSOs on our candidate list.  This analysis could indicate
that our list of surviving candidate YSOs is substantially
contaminated, or it could reflect instead the fact that all the
searches to date (including our prior Spitzer survey) have focused on
regions of high \av, and the ``discovery space'' is in the regions of
lower \av.  

Figure~\ref{fig:avimage} shows the distribution of recovered YSOs, new
YSO candidates, and rejected YSO candidates on the \av\ map from
Froebrich \etal\ (2007). As inferred above, the rejected objects tend
to be evenly distributed, and the recovered YSOs tend to be clustered
in regions of high \av. The new objects are not particularly
clustered, but not evenly distributed either. There is an apparent
clump of objects in the northwest, towards Perseus (see also
Fig.~\ref{fig:whereiras100}); most of these objects have excesses only
at 22 \mum, which, as discussed below, may be a result of source
confusion.

\subsection{W1 Brightnesses}

\begin{figure*}[tbp]
\epsscale{1}
\plotone{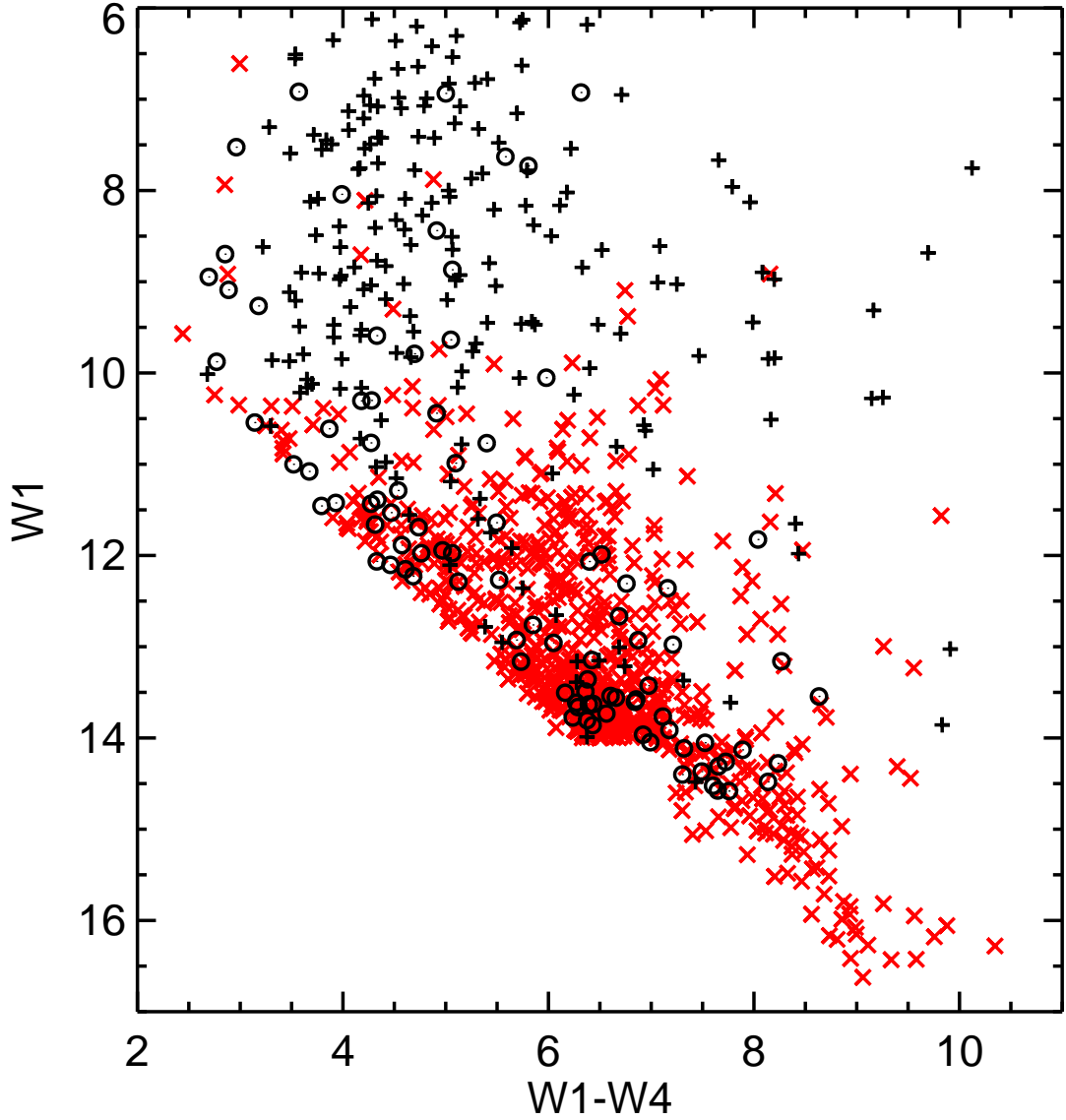}
\caption{The W1 vs.\ W1$-$W4 ([3.4] vs.\ [3.4]$-$[22]) color-magnitude 
diagram for the recovered previously-identified YSOs (+), the rejected
candidates ($\times$), and the best remaining YSO candidates
(circles). }
\label{fig:w1w4}
\end{figure*}

The color cuts in Koenig \etal\ (2011) are performed in, among other
color spaces, W1 vs.\ W1$-$W4 ([3.4] vs.\ [3.4]$-$[22]). This diagram,
and its Spitzer analog ([3.6] vs.\ [3.6]$-$[24]), have proven to be
very useful in identifying YSO candidates (see, e.g., Rebull \etal\
2010).  Figure~\ref{fig:w1w4} shows this diagram for the 94 YSO
candidates, rejected candidates, and recovered YSOs. Photospheres
(stars without disks) would have a W1$-$W4 color near zero, and
galaxies near W1$\sim$16, W1$-$W4$\sim$7, but they have already been
removed by the Koenig \etal\ (2011) process. Most of the
previously-identified YSOs are bright, and most of the contaminants
are faint. The new YSO candidates span the range of bright and faint.

\begin{figure*}[tbp]
\epsscale{1}
\plotone{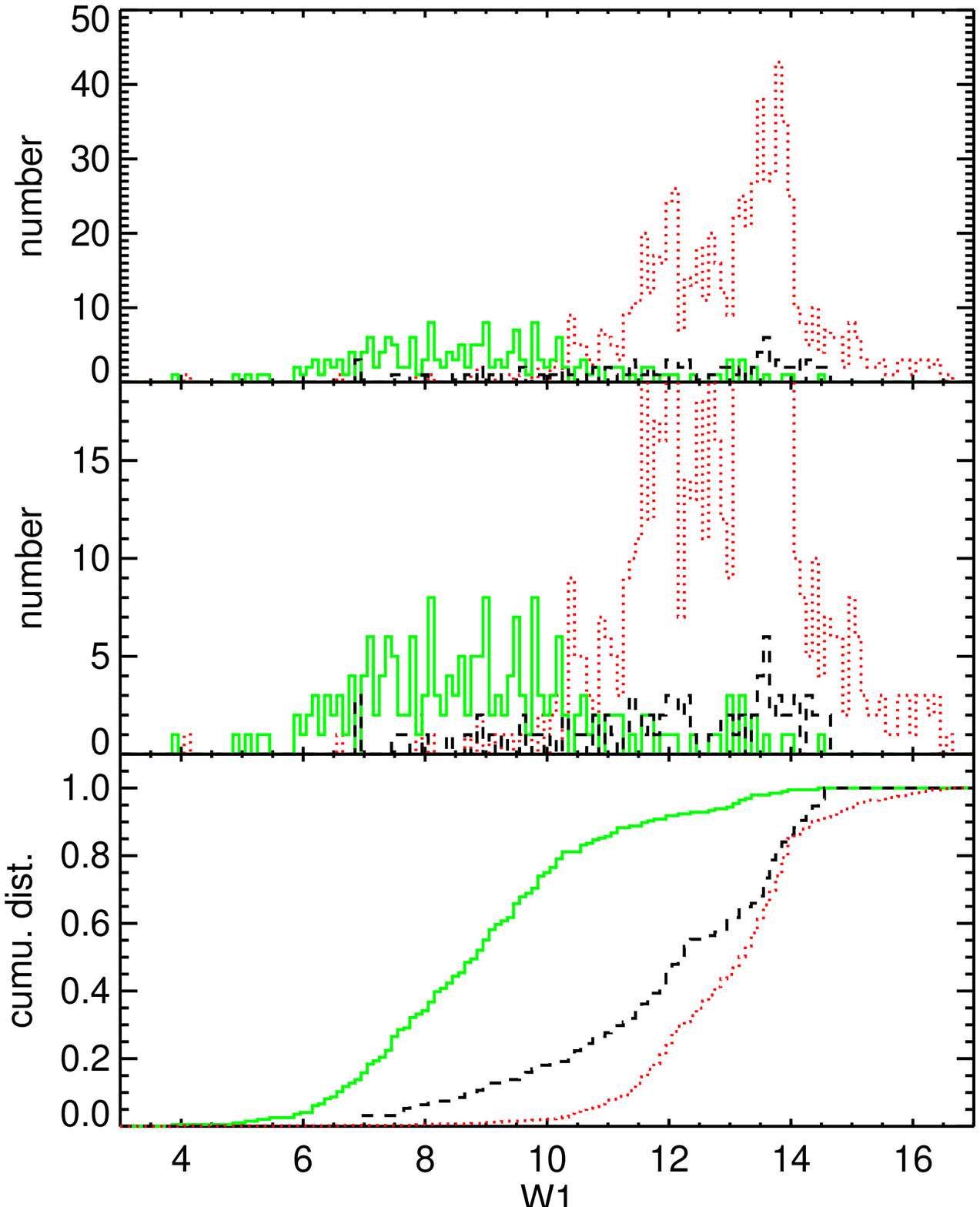}
\caption{Histogram of W1 (3.4 \mum\ values; top two, where middle is
zoomed-in version of the top), and cumulative distribution of W1 
values (bottom), for recovered known YSOs (green, solid line),
rejected contaminants (red, dotted line), and new YSO candidates
(black, dashed line).  Unlike the \av\ distributions in the previous
figure, here the distribution of values for the candidates much more
similar to that for the previously known YSOs than that for the
contaminants.}
\label{fig:w1}
\end{figure*}

The distribution of brightnesses at W1 (3.4 \mum) can be seen
explicitly in Figure~\ref{fig:w1}.   Figure~\ref{fig:w1} suggests a
different conclusion than Figure~\ref{fig:av}; here, the new
candidates are intermediate in cumulative brightness distribution
between the previously known YSOs (brighter) and the contaminants
(fainter). This assessment is reinforced by the formal values
calculated for 2-sided 2-dimensional Komolgorov-Smirnov (KS) tests.
This is consistent with broad expectations; most of the background
galaxies will be faint, most of the surveys to date will have found
the bright YSOs, and the ``discovery space'' for new YSOs is on the
faint end of the distribution of expected brightnesses for YSOs in
Taurus.    Unextincted YSOs in Taurus above the hydrogen burning limit
should be no fainter than 13th magnitude at W1. Thus, fainter YSOs
should show evidence of extinction, and much fainter ones could only
be edge-on disks or more distant emission-line objects.

Figure~\ref{fig:w1} suggests that our YSO list is contaminated to a
larger degree at the faint end of the distribution of 3.4 \mum\
magnitudes. There is a small peak in the distribution of candidate
magnitudes at the same location of the highest peak of contaminant
magnitudes (W1$\sim$13.5 mag), suggesting that the highest fraction of
contaminants per bin will be found in those bins. Similar results are
obtained for, e.g, a histogram of $H$ magnitudes. However, it is worth
noting that in our earlier studies of Taurus, we found a large number
of bright background stars (e.g., AGBs) that appeared to have
excesses, but are not members of Taurus.  


\subsection{Near-IR colors}

\begin{figure*}[tbp]
\epsscale{1}
\plotone{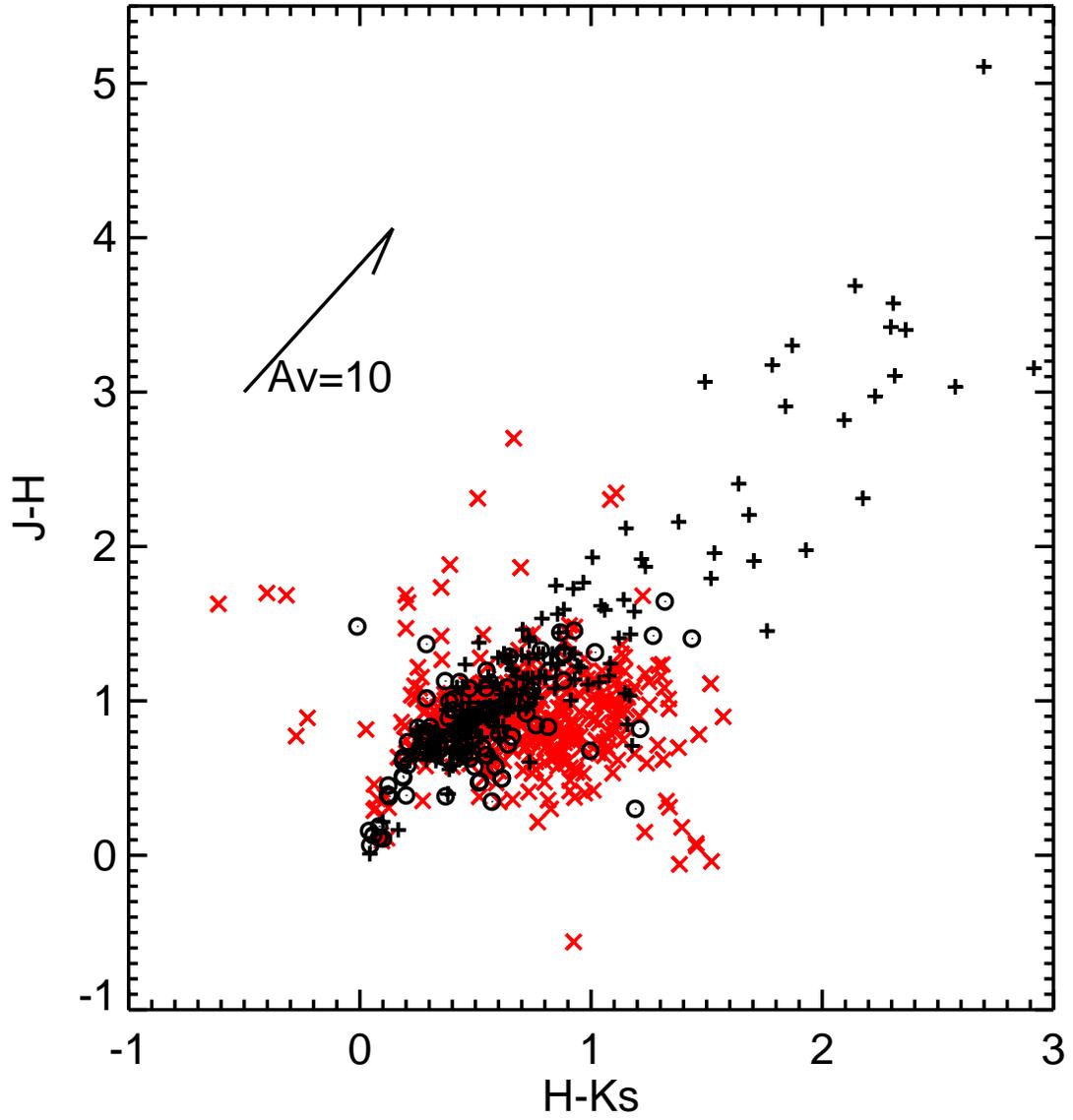}
\caption{The $JHK_s$ color-color diagram for the recovered
previously-identified YSOs (+), the rejected candidates ($\times$),
and the best remaining YSO candidates (circles). }
\label{fig:jhk}
\end{figure*}

Figure~\ref{fig:jhk} shows the $JHK_s$ color-color diagram for the 94
YSO candidates, rejected candidates, and recovered known YSOs.
Consistent with expectations and with Figure~\ref{fig:av} above, the
distribution  of the previously-identified YSOs extends from the locus
of normal young and accreting stars with near-infrared excesses in the
direction of the reddening vector. Encouragingly, the distribution of
contaminants is broad and generally consistent with non-stellar
soures. Few of the YSO candidates have large values of \av, consistent
with Figure~\ref{fig:av}.  There are 11 objects with $(H-K_s)>0.75$
and $(J-H)>1$ : J040204.41+251210.3, J040805.16+300627.6,
J041419.20+220138.0, J043131.61+293819.0, J043327.90+175843.8,
J044512.68+194501.1, J044719.33+325449.8, J044801.88+294902.1,
J044813.48+292453.5, J045352.77+222813.7, and J045913.28+320201.3.
These objects are in the right regime of the $JHK_s$ diagram to be
higher-quality YSO candidates. However, as can be seen in the
Appendix, we have no ancillary photometric data for most of these, and
for the two where we do have such data, the WISE measurements do not
agree with those from Akari (J044813.48+292453.5) or IRAC
(J043131.61+293819.0=0431316+293818 in our Spitzer catalog).  This is
not generally the case -- usually the WISE, Akari, and Spitzer
measurements are in good agreement. One reason they might not agree
would be the differing resolution of the instruments -- if there is
source confusion or the source is resolved (e.g., it is a likely galaxy),
then the photometry would not agree. However, these objects are point
sources in all of the images we examined.  Another reason why the
photometry might be different is that the young stars themselves may
vary between epochs (see, e.g., Morales-Calderon \etal\ 2011, or
Rebull 2010 and references therein). Additional observations are
warranted.

\subsection{Optical data}

The SDSS data do not extend over our entire studied region, but
optical data can be very helpful in ruling out YSO candidates.  There
are matches to SDSS spectra for only 27 objects in our entire set, 15
of which are tagged galaxies (including known YSOs CoKu Tau/1 and GV
Tau; presumably the SDSS pipeline is confused by emission lines in
these sources).  Eleven more of the SDSS spectra are for previously
identified YSOs. Thirteen of the objects that we have placed on our
``reject'' list have SDSS spectra, all of which are galaxies. Just one
of our surviving YSO candidates has a SDSS spectrum, and it is
identified as an M5. Since our placement of objects on the reject or
candidate YSO list was independent of the SDSS spectra, this is
encouraging, and suggests that our classification of the remaining
objects without SDSS spectra may be correct on average.

\begin{figure*}[tbp]
\epsscale{1}
\plotone{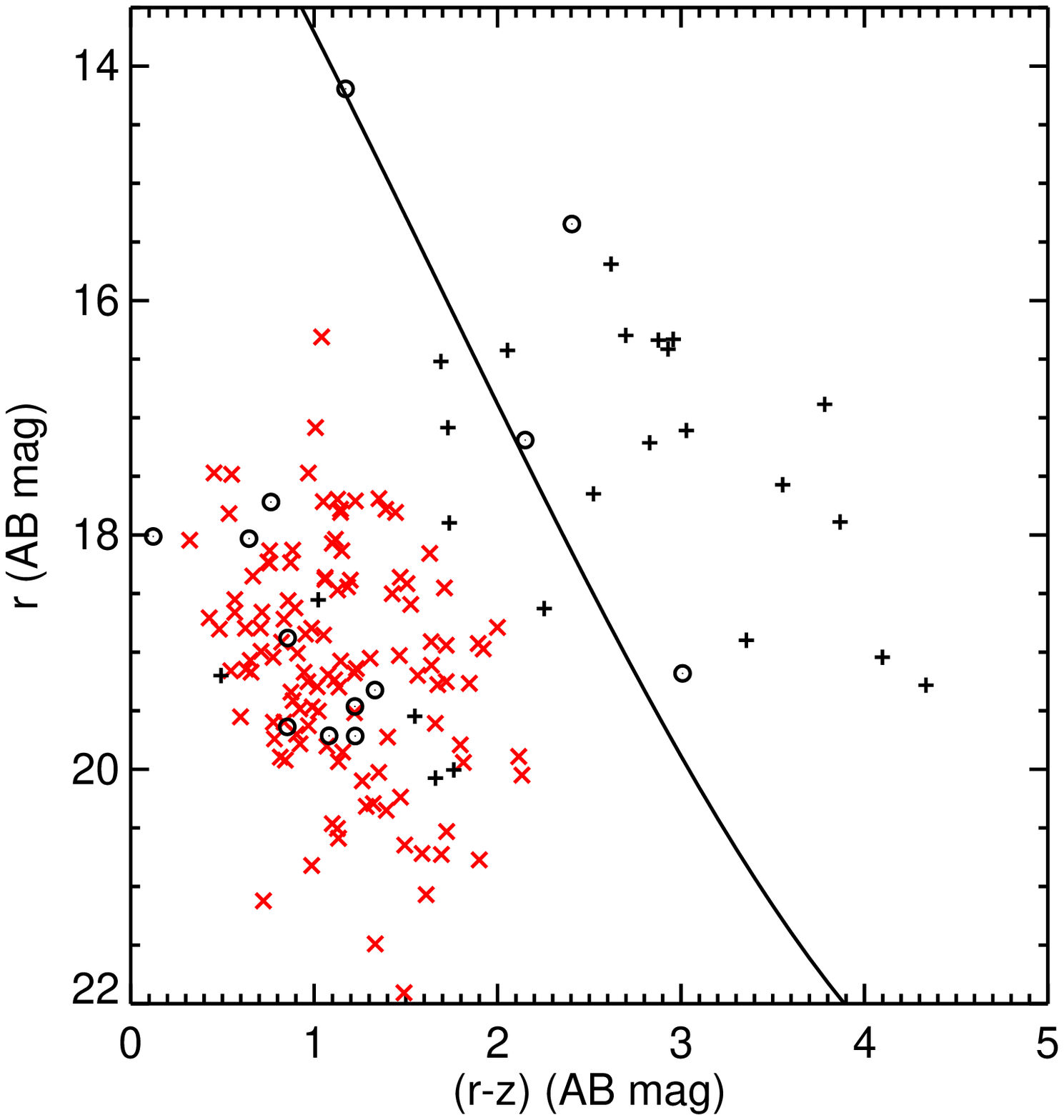}
\caption{SDSS $r$ vs.\ $(r-z)$ in AB magnitudes for the recovered
previously-identified YSOs (+), the rejected candidates ($\times$),
and the best remaining candidates (circles).  The main sequence from
Bochanski \etal\ (2010) is the solid line. Legitimate Taurus objects
could be below the main sequence as a result of edge on disks; in
those cases, we are seeing scattered light from substantial infrared
excesses viewed at a high inclination angle.}
\label{fig:sdss}
\end{figure*}

There are SDSS $z$ measurements for $\sim$26\% of our entire set, but
only $\sim$15\% of our surviving YSO candidates list.
Figure~\ref{fig:sdss} shows the $r$ vs.\ $(r-z)$ color-magnitude
diagram for our objects. Most of the recovered known YSOs are in the
expected location, e.g., to the red side in the diagram.  Just four of
our candidate YSOs are in the main portion of the distribution of
known YSOs (J041240.69+243815.6, J041551.98+315514.0,
J042625.87+351507.6, and J045808.02+251852.6).  It is worth noting
that the distribution of previously identified YSOs as well as
candidates extends into the regime occupied by the contaminants on the
blue side, but that all five of the previously identified YSOs in that
area ($r<$1.9 and $(r-z)<$1.9) are actually unconfirmed YSOs from
Rebull \etal\ (2010) (J041604.83+261800.9, J041810.60+284447.0,
J041823.20+251928.0, J041858.99+255740.0, and J041940.48+270100.7),
and may in the end turn out to not be Taurus members. The nine new YSO
candidates in that regime may have a similar fate; they are 
J041141.34+261341.3, J041831.61+352732.7, J042146.44+240147.1,
J042220.80+170812.0, J042659.19+183548.7, J043326.74+204758.7,
J044428.98+201537.5, J045522.17+250923.2, and
J045943.63+252647.2.

\subsection{Size of Infrared Excess}

\begin{figure*}[tbp]
\epsscale{1}
\plotone{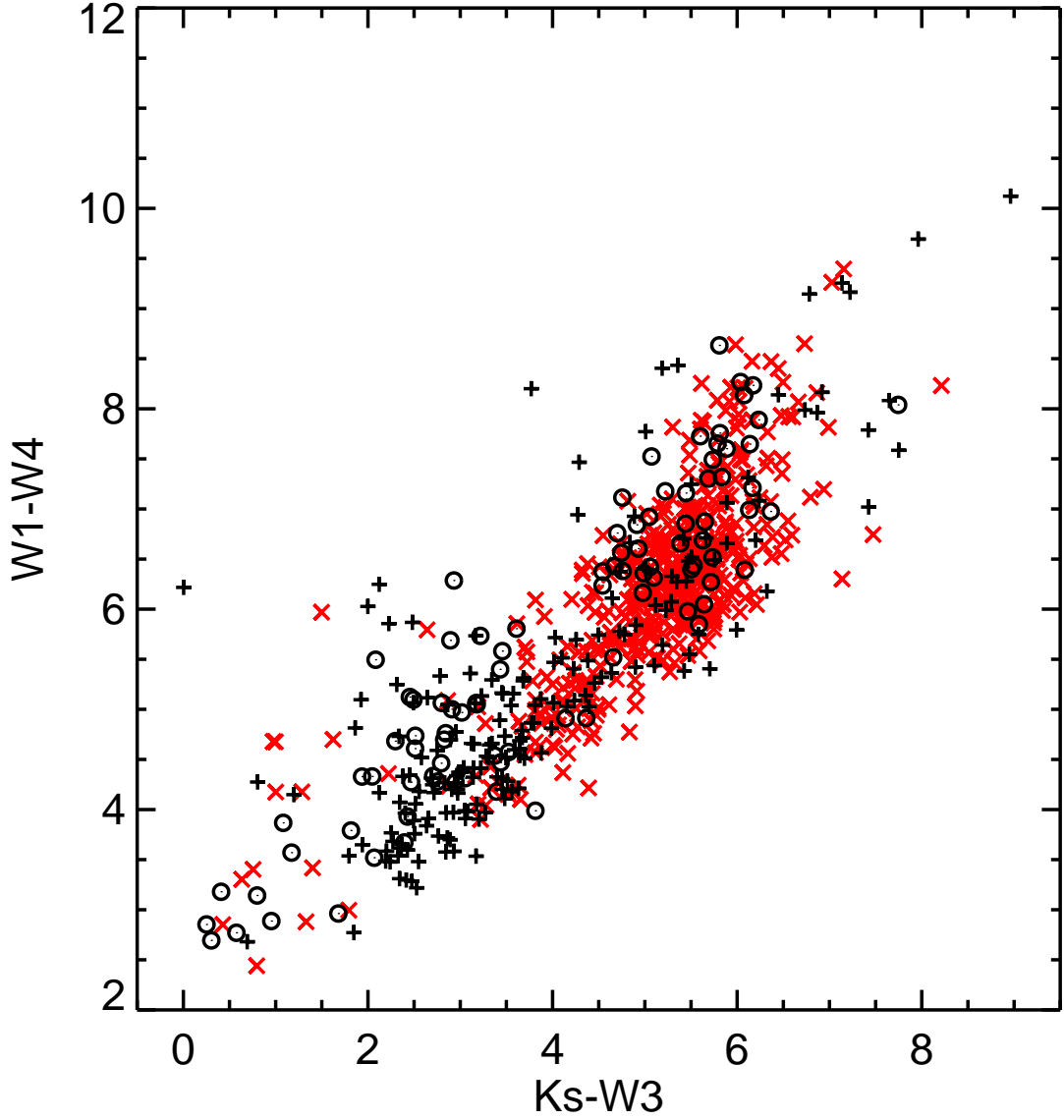}
\caption{W1-W4 ([3.4]$-$[22]) vs.\ \ks$-$W3 (\ks$-$[12]) for the
recovered previously-identified YSOs (+), the rejected candidates
($\times$), and the best remaining candidates (circles).}
\label{fig:excesssize}
\end{figure*}

Since the WISE spatial resolution is relatively low, especially if
there is an infrared excess seen only at 22 \mum, one must be
suspicious of source confusion.  The long wavelength excess objects
may also be the most interesting astrophysically, since they could
have large inner disk holes.  We have investigated each of the images,
and have not been able to identify source confusion. Objects that we
have identified that have excesses only at 22 \mum\ can be selected
out of Figure~\ref{fig:excesssize}. Objects with small \ks$-$W3
excesses ($<$2) generally also have small W1$-$W4 excesses, but the
eleven that also have W1$-$W4$<$4 are the ones for which we are most
suspicious of source confusion:  J035958.81+312901.1,
J040159.15+321941.2, J040651.36+254128.3, J041025.61+315150.7,
J041752.27+360720.2, J041946.18+194539.8, J042954.77+291228.5,
J045808.02+251852.6, J050357.98+265828.8, and J050450.13+264314.6.

\subsection{Distribution of SED Classifications}

In Rebull \etal\ (2010), we compared a variety of YSO selection
methods, and we also had enough ancillary data that we were able to
categorize most of the objects that WISE sees in this overlap region;
Figure~\ref{fig:location} shows relatively few WISE-identified new
YSO candidates in the region where we had Spitzer data (compare to
Figure~\ref{fig:where}).  In the region of Spitzer coverage, we can
compare the selection statistics for the Koenig \etal\ (2011) method
with that from Gutermuth \etal\ (2008, 2009). There are IRAC-1
measurements for $\sim$25\% of our set.  There are 210 for which we
can calculate a Gutermuth \etal\ classification, 119 of which are
recovered known YSOs. Both the Koenig \etal\ and Gutermuth \etal\
methods return SED classifications of objects.
Table~\ref{tab:comparemeth} shows a comparison of the methods. For the
most part, the classifications agree -- Gutermuth \etal\ identify
relatively few of the sources as contaminants (9\%), and both methods
agree that 31\% of the sample are Class I and 40\% are Class II.

Finally, we consider, among the objects we have recovered, rejected,
or retained as YSO candidates, the distribution of classifications
returned by the Koenig \etal\ (2011) method; see
Table~\ref{tab:xkclass}.  The rejected sources are predominantly
categorized as Class I candidates, consistent with the observation
that many have very flat SEDs suggestive of extragalactic objects (or
a reflection of our selection bias against very flat SEDs). The
recovered YSOs are mostly categorized as Class II, consistent with
what is already known in Taurus. The set of surviving YSO candidates
is also mostly categorized as Class II objects.

\section{Conclusions}
\label{sec:concl}


We have presented the results of our search for new candidate Taurus
members using 2MASS+WISE data over $\sim$260 square degrees.  We used
NIR+MIR colors to select a set of YSO candidates following methodology
presented by Koenig \etal\ (2011).  We compared the resultant list of
YSO candidates to our own catalog from Rebull \etal\ (2010), updated
to include Taurus members not in our original Spitzer map, the 2MASS
Extended Source Catalog, several SDSS stripes (photometry plus
spectroscopy), and Akari data.  We recover 196 previously-identified
young stars in the direction of Taurus with infrared excess, and we
identify 686 objects that are confirmed or likely galaxies, 13
foreground or background stars, 1 planetary nebula, 24 objects likely
subject to confusion based on inspection of the images, and 94 new YSO
candidates.  As was the case for our smaller Spitzer-based search for
YSO candidates (Rebull \etal\ 2010), supporting ancillary data and
manual inspection of the images and SEDs for each YSO candidate are
critical. The YSO candidates are broadly distributed in (projected)
space, unlike the previously-identified YSOs, which are generally
clustered. Few are in regions of high \av\ (again, unlike the
recovered YSO sample), but they are generally fainter in 3.4 \mum\
than the previously-identified sample.   There is likely to be
contamination lingering in this list.  We identify a few objects that
are particularly likely to be legitimate YSOs, and a few that are
particularly likely to be contaminants. Follow-up spectra to obtain
spectral types and to search for emission lines and other indications
of youth are warranted.

\begin{deluxetable}{ccccccccccccccccc}
\tablecaption{Multiband measurements of WISE-identified YSOs with
previous indications of youth in the literature towards the Taurus
region.\tablenotemark{a}\label{tab:knownysos}}
\rotate
\tabletypesize{\tiny}
\tablewidth{0pt}
\tablehead{
\colhead{line} & \colhead{catalog number}  & 
\colhead{name}  &
\colhead{W1 (3.4 \mum) } & \colhead{W2 (4.6 \mum) } & 
\colhead{W3 (12 \mum) } & 
\colhead{W4 (22 \mum) } & \colhead{notes} \\
\colhead{num.} &  & & \colhead{(mag)} & \colhead{(mag)}&
\colhead{(mag)}& \colhead{(mag)} }
\startdata
  1& J040443.06+261856.3&           IRAS04016+26102&       7.67$\pm$       0.03&       5.16$\pm$       0.03&       2.91$\pm$       0.02&       0.01$\pm$       0.02& \\
  2& J041339.46+292113.8&                      none&      12.78$\pm$       0.03&      11.74$\pm$       0.02&       9.43$\pm$       0.04&       7.40$\pm$       0.11& still a candidate YSO\tablenotemark{b}  \\
  3& J041353.28+281123.1&          IRAS04108+2803 A&       9.44$\pm$       0.02&       8.28$\pm$       0.02&       5.47$\pm$       0.02&       3.60$\pm$       0.02& \\
  4& J041354.70+281132.6&          IRAS04108+2803 B&       9.44$\pm$       0.02&       7.36$\pm$       0.02&       4.33$\pm$       0.02&       1.46$\pm$       0.01& \\
  5& J041357.37+291819.1&            IRAS04108+2910&       8.00$\pm$       0.02&       6.98$\pm$       0.02&       4.96$\pm$       0.02&       2.97$\pm$       0.02& \\
  6& J041412.29+280837.1&           IRAS04111+2800G&      13.86$\pm$       0.03&      11.73$\pm$       0.03&       8.97$\pm$       0.03&       4.03$\pm$       0.02& \\
  7& J041412.91+281212.3&              V773 Tau ABC&       6.64$\pm$       0.02&       5.87$\pm$       0.02&       3.87$\pm$       0.02&       1.91$\pm$       0.01& \\
  8& J041413.58+281249.0&                    FM Tau&       8.07$\pm$       0.02&       7.48$\pm$       0.02&       5.21$\pm$       0.02&       3.03$\pm$       0.03& \\
\enddata
\tablenotetext{a}{See full table description in
Section~\ref{sec:knownysos}.}
\tablenotetext{b}{Object identified in Rebull \etal\ (2010) as a
candidate Taurus member, but discussed there as needing additional
data to confirm/refute it as a Taurus member.}
\end{deluxetable}

\begin{deluxetable}{ccccccccccccccc}
\tablecaption{Multiband measurements of WISE-identified rejected YSO
candidates in Taurus region\label{tab:rejysos}}
\rotate
\tabletypesize{\tiny}
\tablewidth{0pt}
\tablehead{
\colhead{line} & \colhead{catalog number}  & 
\colhead{name}  &
\colhead{W1 (3.4 \mum) } & \colhead{W2 (4.6 \mum) } & 
\colhead{W3 (12 \mum) } & 
\colhead{W4 (22 \mum) } & \colhead{notes} \\
\colhead{num.} &  &  & \colhead{(mag)} & \colhead{(mag)}&
\colhead{(mag)}& \colhead{(mag)} }
\startdata
  1& J035135.81+260914.3&     2MASX03513585+2609139&      13.82$\pm$       0.03&      12.94$\pm$       0.03&       9.67$\pm$       0.05&       7.47$\pm$       0.13& galaxy: extended in 2MASS, SED shape \\
  2& J035228.38+252002.4&                      none&      12.14$\pm$       0.03&      10.94$\pm$       0.02&       8.36$\pm$       0.03&       6.34$\pm$       0.06& galaxy: SED shape \\
  3& J035234.58+252729.5&                      none&      13.37$\pm$       0.03&      12.28$\pm$       0.02&       9.19$\pm$       0.04&       6.54$\pm$       0.07& galaxy: SED shape \\
  4& J035300.20+274013.1&                      none&      13.89$\pm$       0.03&      12.88$\pm$       0.03&       9.99$\pm$       0.06&       7.81$\pm$       0.16& galaxy: SED shape \\
  5& J035300.52+290542.9&                      none&      12.05$\pm$       0.02&      11.78$\pm$       0.03&       8.40$\pm$       0.03&       7.03$\pm$       0.08& galaxy: SED shape \\
  6& J035309.58+272953.5&     2MASX03530961+2729541&      12.27$\pm$       0.03&      11.11$\pm$       0.02&       8.36$\pm$       0.03&       6.23$\pm$       0.05& galaxy: extended in 2MASS, SED shape \\
  7& J035322.85+281358.1&                      none&      14.77$\pm$       0.04&      13.13$\pm$       0.03&       9.69$\pm$       0.05&       6.96$\pm$       0.08& galaxy: SED shape \\
  8& J035334.67+270854.1&                      none&      12.78$\pm$       0.03&      11.67$\pm$       0.02&       9.02$\pm$       0.03&       6.63$\pm$       0.06& galaxy: SED shape \\
\enddata
\end{deluxetable}

\begin{deluxetable}{cccccccccccccc}
\tablecaption{Multiband measurements of WISE-identified new YSO
candidates in Taurus region\label{tab:newysos}}
\rotate
\tabletypesize{\tiny}
\tablewidth{0pt}
\tablehead{
\colhead{line} & \colhead{catalog number}  & 
\colhead{name}  &
\colhead{W1 (3.4 \mum) } & \colhead{W2 (4.6 \mum) } & 
\colhead{W3 (12 \mum) } & 
\colhead{W4 (22 \mum) } & \colhead{notes} \\
\colhead{num.} & & & \colhead{(mag)} & \colhead{(mag)}&
\colhead{(mag)}& \colhead{(mag)} }
\startdata
  1& J035323.82+271838.3&                      none&      11.29$\pm$       0.03&      10.58$\pm$       0.02&       8.66$\pm$       0.03&       6.75$\pm$       0.08& within Pleiades tidal radius (sec.~4.1). \\
  2& J035412.25+312034.0&                      none&      12.10$\pm$       0.03&      11.76$\pm$       0.02&       9.69$\pm$       0.05&       7.64$\pm$       0.14& \\
  3& J035454.27+295944.0&                      none&       8.04$\pm$       0.02&       7.34$\pm$       0.02&       5.31$\pm$       0.02&       4.05$\pm$       0.02& \\
  4& J035519.09+233501.8&                      none&      13.60$\pm$       0.03&      12.79$\pm$       0.03&       9.55$\pm$       0.05&       6.76$\pm$       0.07& within Pleiades tidal radius (sec.~4.1). \\
  5& J035535.05+291319.3&                      none&      13.86$\pm$       0.03&      13.06$\pm$       0.03&      10.19$\pm$       0.06&       7.43$\pm$       0.13& within Pleiades tidal radius (sec.~4.1). \\
  6& J035919.63+310730.1&                      none&      13.56$\pm$       0.03&      12.39$\pm$       0.03&       9.08$\pm$       0.03&       6.90$\pm$       0.08& \\
  7& J035935.02+315409.6&     2MASX03593501+3154090&      13.55$\pm$       0.03&      13.12$\pm$       0.03&       8.40$\pm$       0.03&       4.91$\pm$       0.03& \\
  8& J035958.81+312901.1&                  HD281395&       8.95$\pm$       0.02&       8.95$\pm$       0.02&       8.69$\pm$       0.03&       6.25$\pm$       0.05& excess just at 22 (sec.~4.6).  \\
\enddata
\end{deluxetable}

\begin{deluxetable}{cccc}
\tablecaption{Comparison of Gutermuth \etal\ (2008, 2009) and Koenig
\etal\ (2011) methods \label{tab:comparemeth}}
\tablewidth{0pt}
\tablehead{
\colhead{Koenig \etal\ classification}  & 
\colhead{Gutermuth \etal\ :}  &
\colhead{Gutermuth \etal\ :} & \colhead{Gutermuth \etal\ :}\\
\colhead{} & \colhead{contaminant} & \colhead{Class 0/I} & \colhead{Class II} }
\startdata
Class I         &   2  &  65   &   19 \\
Class II        &   5  &  17   &   84 \\
Transition Disk &  12  &  2    &   4  \\
\enddata
\end{deluxetable}

\begin{deluxetable}{cccc}
\tablecaption{Breakdown of the Koenig
\etal\ (2011) classifications \label{tab:xkclass}}
\tablewidth{0pt}
\tablehead{
\colhead{sample}  & 
\colhead{Class I }  &
\colhead{Class II} & \colhead{Transition Disk}}
\startdata
of the rejects (number)    &  365 &     211 &      148\\
of the rejects (fraction)  & 0.50 &    0.29 &  0.20\\
of the recovered (number)  &   62 &     122 &       12\\
of the recovered (fraction)& 0.32 &    0.62 & 0.06\\
of the candidates   (number)    &   28 &      46 &       20\\
of the candidates   (fraction)  & 0.29 &    0.49 &  0.21\\
\enddata
\end{deluxetable}

\clearpage

\appendix

\section{SEDs for all of the candidates}

For each of our new candidate Taurus members, we provide an SED here
in this on-line-only Appendix. Notation is as follows: $+$--literature
Johnson photometry, *--Sloan photometry, $\times$--CFHT  photometry,
diamonds--2MASS, circles--IRAC, stars--WISE, triangles--Akari,
squares--MIPS. Limits for any  band, if available, are indicated by
arrows. The wavelength is in microns; $\lambda F_{\lambda}$ is  in cgs
units (erg s$^{-1}$ cm$^{-2}$). Each plot has the WISE catalog
number, which can also be found in Table~\ref{tab:newysos}.

\acknowledgements 

This publication makes use of data products from the Wide-field
Infrared Survey Explorer, which is a joint project of the University
of California, Los Angeles, and the Jet Propulsion
Laboratory/California Institute of Technology, funded by the National
Aeronautics and Space Administration. This research has made use of
NASA's Astrophysics Data System (ADS) Abstract Service, and of the
SIMBAD database, operated at CDS, Strasbourg, France.  This research
has made use of data products from the Two Micron All-Sky Survey
(2MASS), which is a joint project of the University of Massachusetts
and the Infrared Processing and Analysis Center, funded by the
National Aeronautics and Space Administration and the National Science
Foundation.  These data were served by the NASA/IPAC Infrared Science
Archive, which is operated by the Jet Propulsion Laboratory,
California Institute of Technology, under contract with the National
Aeronautics and Space Administration.  This research has made use of
the Digitized Sky Surveys, which were produced at the Space Telescope
Science Institute under U.S. Government grant NAG W-2166. The images
of these surveys are based on photographic data obtained using the
Oschin Schmidt Telescope on Palomar Mountain and the UK Schmidt
Telescope. The plates were processed into the present compressed
digital form with the permission of these institutions.

The research described in this paper was partially carried out at the
Jet Propulsion Laboratory, California Institute of Technology, under
contract with the National Aeronautics and Space Administration.

\end{document}